**Published in Applied Physics A: 2003**

# Structural and Magnetoelectric Properties of $MFe_2O_4$ - PZT (Ì = Ni, Co) and $La_x(Ca, Sr)_{1-x}MnO_3$ - PZT Multilayer Composites

G. Srinivasan and E. T. Rasmussen
Physics Department, Oakland University, Rochester, Michigan 48309-4401, USA

A.A. Bush, K.E. Kamentsev, V.F. Meshcheryakov, and Y.K. Fetisov
Moscow State Institute of Radio Engineering, Electronics and Automation,
117454 Moscow, Russia

## Abstract

Thick film layered magnetoelectric composites consisting of ferromagnetic and ferroelectric phases have been synthesized with nickel ferrite (NFO), cobalt ferrite (CFO), $La_{0.7}Sr_{0.3}MnO_3$ (LSMO), or $La_{0.7}Ca_{0.3}MnO_3$ (LCMO) and lead zirconate titanate (PZT). Structural, magnetic and ferromagnetic resonance characterization shows evidence for defect free ferrites, but deterioration of manganite parameters. The resistivity and dielectric constants are smaller than expected values. The magnetoelectric effect (ME) is stronger in ferrite-PZT than in manganite-PZT. The ME voltage coefficient $\alpha_E$ at room temperature is the highest in NFO-PZT and the smallest for LCMO-PZT. The transverse ME effect is an order of magnitude stronger than the longitudinal effect. The magnitude of $\alpha_E$ correlates well with magnetic permeability for the ferrites.

Email: srinivas@oakland.edu; FAX: 1-248-370-3408

Multilayer composites consisting of alternative layers of magnetostrictive and piezoelectric materials are observed to show a strong magnetoelectric (ME) effect and are of interest for studies on ME interactions and for potential use in devices [1-5]. Such structures when placed in an external ac magnetic field $\delta H$ undergo deformation of the magnetic layer due to magnetostriction, resulting in an electrical polarization of the dielectric layer due to piezoelectric effect [6,7]. The ac electrical field $\delta E$ produced in the structure is a "product-property" and is proportional to the piezomagnetic and piezoelectric coefficients of the constituent phases [8]. The magnetoelectric voltage coefficient $\alpha_E = \delta E/\delta H$ for such composites is in the range 50-5000 mV/Oe cm, depending on the nature and volume of the magnetic and piezoelectric phases [1-4,6,7]. These values far exceed the ME coefficient of 20 mV/Oe cm for $Cr_2O_3$, the best single phase ME material [9].

Theoretical models for a bilayer with parameters of nickel or cobalt ferrite and lead zirconate titanate predict coefficients as high as 1.5 V/Oe cm [7]. Although our studies reveal strong ME interactions in nickel ferrite – PZT samples, systems such as cobalt ferrite-PZT show a weak coupling [10]. The reason could be deterioration of physical properties of individual layers and poor mechanical coupling between the two phases due to defects at the interface formed during the sample processing [3,4,6]. A detailed and comprehensive investigation of structural, electrical and magnetic properties of the structures is important to elucidate the cause of magnetoelectric coefficient limitations.

This report details the fabrication and characterization of multilayer thick film structures consisting of piezoelectric (lead zirconate titanate) and ferromagnetic (nickel ferrite, cobalt ferrite, and lanthanum-strontium or calcium manganite) phases. The choice of materials for the structures is due to the following reasons. All the materials, except the manganites, have ferromagnetic or ferroelectric phase transition temperatures well above the room temperature. Piezoelectric lead zirconate titanate (PZT) has high dielectric permeability and high piezoelectric coupling constants. Both nickel and cobalt ferrites have high magnetostriction that gives rise to a strong pseudo-piezomagnetic effect in an ac magnetic field. Manganites demonstrate a large magnetostriction, their perovskite crystal structure is likely to result in strong interface bonding in a composite with PZT. Besides, ferromagnets with metallic conductivity are a key ingredient for realizing strong magnetoelectric effect in multilayer structures [7].

Layered samples were prepared with PZT for the piezoelectric layer and nickel ferrite $NiFe_2O_4$ (NFO), cobalt ferrite $CoFe_2O_4$ (CFO),

lanthanum strontium manganite $La_{0.7}Sr_{0.3}MnO_3$ (LSMO), or lanthanum calcium manganite $La_{0.7}Ca_{0.3}MnO_3$ (LCMO) for the magnetostrictive layers. X-ray diffraction studies confirmed single-phase nature of the two oxides. Electrical resistance and dielectric constant for the samples are smaller than expected values. Magnetization for the ferrite-PZT samples is in agreement with bulk values, but is smaller than expected values for manganite-PZT. Ferromagnetic resonance measurements revealed a large anisotropy in manganite – PZT and CFO – PZT samples. Magnetoelectric studies indicate the strongest ME coupling in NFO – PZT layered composites and the weakest coupling in LCMO – PZT samples. We relate the strength of the ME coupling to piezomagnetic coupling and magnetic and dielectric parameters for the samples. The composites have the potential to form the basis for the development of a new class of magnetic sensors and transducers for use in solid state microelectronics and microwave devices.

## 2. Samples fabrication

Multilayer structures consisting of alternate layers of ferroelectric and ferromagnetic oxides were fabricated from thick films synthesized by the doctor blade technique [11]. The fabrication process contained the following main steps: a) preparation of submicron size powders of constituent oxides; b) preparation of 10-40 μm thick films tapes by doctor blade techniques; and c) lamination and sintering of multilayers. The ferrite and manganite powders were obtained by standard ceramic techniques. Commercial PZT powder was used [12]. Thick films were made from casts prepared by mixing powders with a solvent (ethyl alcohol), plasticizer (butyl benzyl phthalate), and a binder (polyvinyl butyral) in a ball milling for 24 hrs. The slurries were cast on mylar sheets into tapes of 10-40 μm in thickness with a tape caster. The thick film preparation by tape casting is described in detail in Ref.11. The films were removed from the mylar substrate and arranged to obtain the desired structure, laminated under high pressure (3000 psi) and high temperature (400 K), and sintered at 1200-1500 K.

Using the above procedure 1-5 $cm^2$ multilayer structures were fabricated. Samples were made with a layer thickness of 11-35 μm. For all the samples, the thickness of magnetic oxides and PZT were kept the same. The number of piezoelectric layers was $n$ = 8-20 and the number of ferromagnetic layers was $n$+1. The surface morphology and the cross-section of the samples were examined with a metallurgical microscope. Samples contained fine grains (1-5 μm) and some open pores. The porosity ranged from 5 to 8% depending on the samples processing temperature. The cross-section studies showed well-bonded structure with uniform thickness for the magnetic phase and PZT. To pole the PZT layers of the samples, they were heated up to 420 K and then cooled in an external electrical field of 20-50 kV/cm applied perpendicular to the sample plane. The composites studied, their composition, number of layers, thickness of each layer, and measured lattice parameters are summarized in Table 1.

Table 1: Structural parameters of the multilayer samples prepared from thick films obtained by tape casting techniques.

| Sample | Composition of the structure (number of layers) | Sample dimensions ($mm^3$) and layer thickness (μm) | Lattice Parameter (nm) |
|---|---|---|---|
| NFO-PZT | Nickel Ferrite(21)-PZT1(20) | 6 õ 3 õ 0.45<br>11 | $NiFeO_4$ – cubic, 0.8332(2) |
| CFO-PZT | Cobalt Ferrite(10)-PZT(9) | 5 õ 4 õ 0.49<br>26 | $CoFe_2O_4$ – cubic, 0.8375(3) |
| LSMO-PZT | Lanthanum strontium manganite(21)- PZT(20) | 6 õ 4 õ 0.41<br>10 | $(La,Sr)MnO_3$ – pseudo-cubic, 0.3895 nm |
| LCMO-PZT | Lanthanum calcium manganite(9)- PZT(8) | 6 õ 3 õ 0.59<br>35 | $(La,Ca)MnO_3$ – pseudo-cubic, 0.3853 nm |

## 3. Results and Discussion

Multilayer samples of NFO-PZT, CFO-PZT, LSMO-PZT and LCMO-PZT were characterized in terms of structural, electric, magnetic and magnetoelectric parameters. Structural studies were done by x-ray diffraction on multilayers and powdered layered samples. Longitudinal and transverse electrical resistance and dielectric constant were measured for electrical characterization. The composites were then poled in an electric field and piezoelectric and magnetoelectric coupling were measured. The strength of ME coupling was measured for transverse and longitudinal field orientations. Magnetostriction was measured on bulk samples of the ferromagnetic phase for information on the piezomagnetic couplings that are related to the strength of ME effects.

### *3.1 X-ray diffraction measurements*

Structural characterization was carried out using an x-ray diffractometer and $CuK_\alpha$ filtered radiation. The diffraction patterns were recorded at a scan velocity of 0.5 deg/min. Sodium chloride single crystals were used as internal reference. The diffraction patterns for all the structures were obtained on powdered multilayers and also from the layered sample. X-ray diffraction patterns for NFO-PZT powder contained two sets of well-defined and narrow peaks. The first set corresponded to the magnetic phase (NFO), while the second set was identified with the piezoelectric phase (PZT). Main peaks of both sets were of nearly the same intensity. The peak intensities and their angular positions were in agreement with diffraction patterns of pure oxides, NFO and PZT. The diffraction patterns obtained from the surface of NFO-PZT samples contained the same two sets of peaks. However, intensity of PZT peaks was 10% of the nominal value due to shielding of internal PZT layers by the external NFO layer. The observed intensities were in agreement with estimates based on linear absorption coefficient in the outer ferrite layer for the $CuK_\alpha$ radiation.

The structural parameters for all the samples were calculated using the x-ray data and are given in Table 1. Both nickel ferrite and cobalt ferrite have a cubic structure with lattice parameters $a$ = 0.8332(2) nm and 0.8375(2) nm, respectively. Layers of LSMO and LCMO have a distorted perovskite structure with lattice parameters $a$ = 0.3895(2) nm and 0.3853(2) nm, respectively. Unit cell parameters of the oxides are in good agreement with values for bulk materials [13-15]. One draws the following inferences from x-ray diffraction studies. (i) The presence of two characteristic sets of reflections implies that the layered structure is conserved during the sample processing. (ii) No new (impurity) phases are formed at the interfaces because of diffusion. (iii) The narrowness of the peaks in the diffraction patterns and the estimated lattice parameters indicate that fabricated films have the same crystal structure as bulk materials.

### *3.2. Electrical measurements*

Measurements of electrical resistance and capacitance were carried out to probe the quality of the composites. The resistance for a current perpendicular to the composite plane $R_\perp$, in-plane resistance $R_{//}$, and capacitance $C$ were measured and are given in Table 2. For measurements of the transverse resistance and capacitance, the sample surfaces were coated with silver paint and the sample was placed between two copper plates connected to a voltage source. For measurements of the in-plane resistance, two contacts were deposited 4 mm apart on the sample surface and the current was passed in the film plane. All measurements were performed at a frequency of 1 kHz and in the temperature range 0-700$^0$C.

**Table 2.** Electrical parameters of the samples at room temperature

| Sample | $R_\perp$, MΩ | $R_{//}$, MΩ | $C$, pF | ε |
|---|---|---|---|---|
| NFO-PZT | 10.38 | 360 | 66 | 91 |
| CFO-PZT | 13.69 | 4340 | 28 | 37 |
| LSMO-PZT | 66.11 | 0.170 | 23 | 22 |
| LCMO-PZT | 39.02 | 0.025 | 42 | 74 |

The resistance $R_\perp$ is the sum of resistance of the layers in series. The effective resistance is determined by the resistance of PZT layers which is a good dielectric (ρ~$10^{12}$ Ω·cm at room temperature), and the expected $R_\perp$ is on the order of $10^4$ MΩ [15]. Measured values in Table 2, however, are in the range 10-66 MΩ. This decrease in the resistance is probably due to higher than expected conductivity of PZT films in comparison with bulk values or due to the presence of "shorts" in the PZT films since the porosity is on the order of 5-8% for theses samples. To verify this possibility, top layers of several samples were removed by polishing. Voids with a diameter of 50-500 μm were observed on polished surfaces of internal PZT layers. Such regions are most likely formed because of nonuniformity of the film thickness and/or local deformation of layers during high temperature sintering. Thus the decrease in the resistance is possibly caused by defects in the structures. The sample resistance $R_{//}$ is a measure of the conductivity of external magnetic layers of the composite. Data in Table 2 indicate that nickel ferrite and cobalt ferrite layers are good dielectrics. The manganite layers have the lowest resistance, as expected.

Measured capacitance of the samples and effective dielectric permeability of the PZT layers calculated from C are given in Table 2. For the estimation of the effective permittivity ***e***, the multilayer structure was considered as capacitors connected in series through conducting magnetic layers. The largest value of ***e*** = 91 is measured for NFO-PZT and is an order of magnitude smaller than the dielectric permeability for bulk PZT [15]. This result demonstrates a necessity and possibility to improve the quality of the PZT layers.

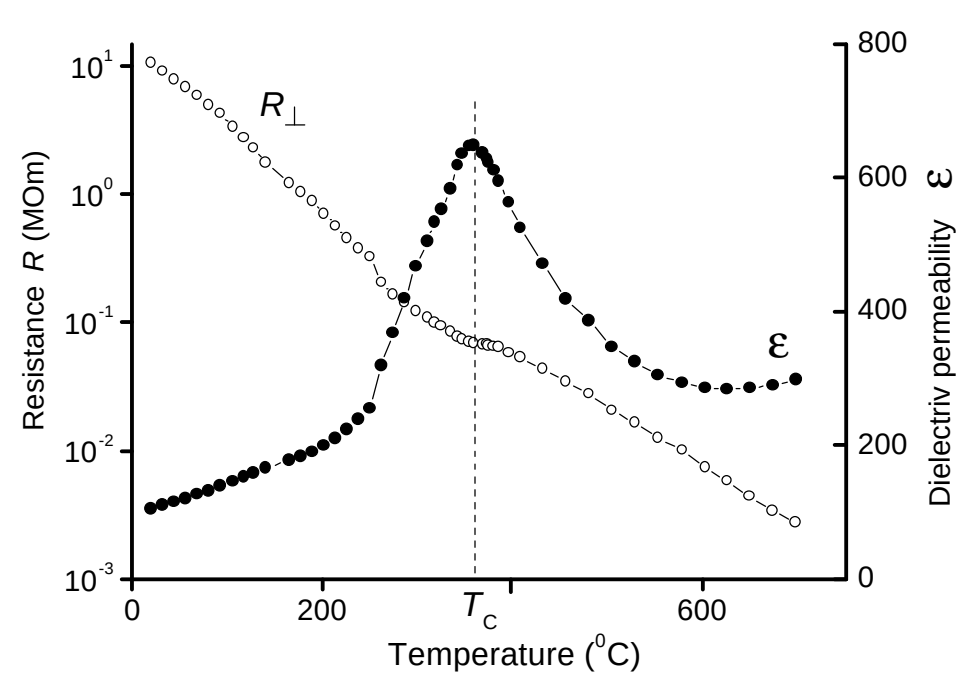

Fig.1: Dependence of the transverse resistance $R_\perp$ and effective dielectric permittivity ***e*** versus temperature for the layered sample of nickel ferrite (NFO)-lead zirconate titanate.

Figure 1 shows measured dependence of the transverse resistance $R_\perp$ and effective dielectric permeability ***e*** vs. temperature for the NFO-PZT sample in the temperature range $T$ = 0-700 $^0$C. Note that $R_\perp(T)$ is in logarithmic scale. The parameter ***e*** ($T$) was calculated from the sample capacitance. There is a local maximum $\boldsymbol{e}_{max}$ ≈ 648 at $T_c$ ~360$^0$Ñ in the dielectric constant. The maximum corresponds to the ferroelectric to paraelectric transition in the PZT layer. The transition temperature agrees well with the phase transition temperature of $T_c$ ~330 $^0$Ñ for PZT [15]. The sample resistance decreased nearly exponentially, by four orders of magnitude in this temperature range, probably because of thermal-activation processes in piezoelectric layers. Note the characteristic discontinuity in $R_\perp(T)$ near $T_c$ ~360 $^0$Ñ in Fig.1.

Thus, the data in Fig.1 show that both the resistance and the dielectric permeability of multilayer structures are one to two orders of magnitude smaller than values for bulk materials. Magnetoelectric studies, discussed later, show that maximum coupling is observed in ferrite-PZT structures where piezoelectric layer has the highest dielectric permeability and the magnetic layer has the lowest resistance.

### *3.3. Magnetic measurements*

Measurements of static magnetization for parallel and perpendicular orientations of external magnetic field with respect to the sample plane were carried out for all the structures in Table 1. A vibrating sample magnetometer was used to obtain magnetization data in fields H up to 6 kOe at room temperature. To record the hysteresis, the sample was first magnetized to saturation and then the magnetization was measured for decreasing H. A similar procedure was employed for data for negative fields. Magnetization of the samples was calculated taking into account volume of the ferromagnetic phase (not for the composite volume).

Figure 2 shows typical magnetization curves for all the samples for both orientations of external magnetic field. Magnetization curves for the samples having different number of layers and layer thickness overlapped each other. One observes in Fig.2(a) for the NFO-PZT sample a strong dependence of M on the orientation of H. For in-plane H, the saturation takes place for at about of 2 kOe, which is much smaller than the saturation field for transverse H. The saturation magnetic moment $M_s$ per unit volume of the NFO phase was 330 G, in agreement with the magnetization of bulk nickel ferrite [13]. No hysteresis was observed for the nickel ferrite-PZT samples. For the CFO-PZT sample (Fig. 2b), there is significant dependence of M on H orientation as expected. Besides, a well-defined hysteresis is observed. Saturation of the sample magnetization occurs for H higher than 6 kOe, where the magnetic moment is equal to 400 G, in agreement with reported values of 425 G [13]. Such an M versus H behavior is indicative of the expected large magnetic anisotropy for cobalt ferrite. Thus the magnetization curves of Fig.2 for ferrite-PZT samples correspond to magnetization for bulk ferromagnetic counterparts. The high temperature processing of multilayers usually leads to (i) an interface stress and (ii) chemical and/or structural inhomogeneities. Any stress is likely to manifest as an anisotropy field. But chemical or structural defects will result in deviation of saturation magnetization from bulk values. Thus the key inference from Fig.2 is that ferrites and PZT cosinter well without the formation of any impurity phases.

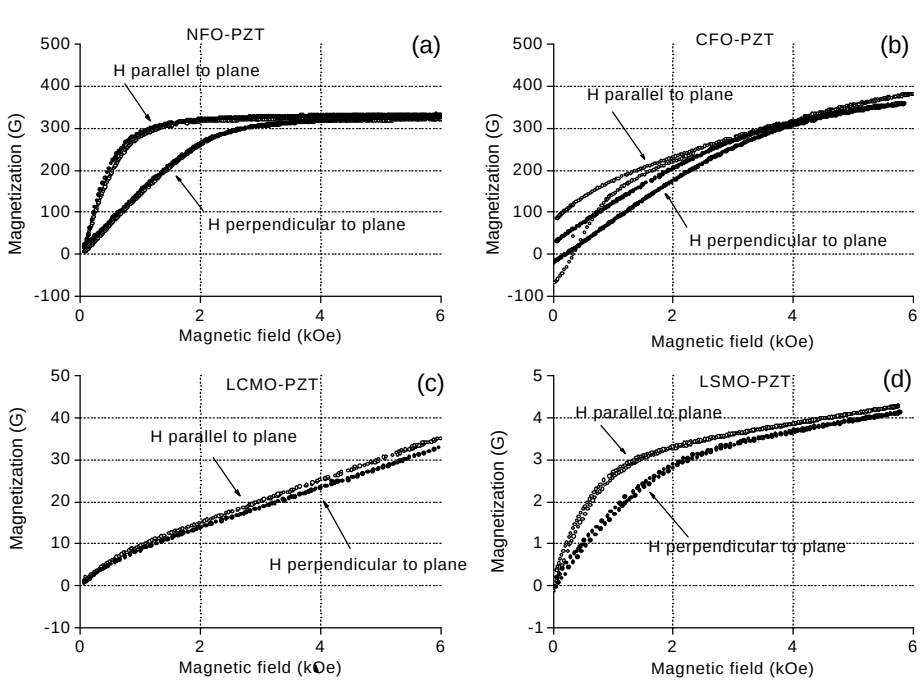

Fig.2: Magnetization as a function of static magnetic field H for the multilayer composite structures at room temperature for (a) NFO-PZT, (b) cobalt ferrite (CFO) – PZT, (c) $La_{0.7}Ca_{0.3}MnO_3$ (LCMO) – PZT and (d) $La_{0.7}Sr_{0.3}MnO_3$ (LSMO) – PZT. The data are for H parallel and perpendicular to the sample plane.

We, however, found evidence for diffusion related degradation of sample parameters for manganite-PZT. Bulk samples of the manganites $La_{0.7}Sr_{0.3}MnO_3$ (LSMO) and $La_{0.7}Ca_{0.3}MnO_3$ (LCMO) order ferromagnetically at 385 K and 260 K, respectively [14]. The composites, however, showed a ferromagnetic $T_c$ of 240 K for LCMO-PZT and 301 K for LSMO-PZT. At room temperature, M is on the order of 40 G for the LCMO-PZT sample (Fig. 2c) and no hysteresis was observed. The magnetization curves for different field orientations overlapped each other since the sample is paramagnetic at room temperature. Results in Fig.2 for LSMO-PZT are somewhat different. The magnetization is relatively small and there is a small difference in magnetization curves corresponding to different orientations of H. There is clear evidence for ferromagnetic ordering at room temperature and a minor hysteresis is observed in Fig.2(d). The saturation magnetization is about of 4 G and is two orders of magnitude smaller than expected values [16]. The reduction in M and $T_c$ agrees with our earlier observation of deterioration of magnetic parameters for the composite when the film thickness is reduced or the sintering temperature is increased [4]. Even though the structural homogeneity for LSMO and PZT is likely to enhance the interface bonding, one also expects diffusion of metal across the interface and a consequent adverse effect on magnetic and dielectric properties of the layered samples. In our earlier study, we measured a substantial reduction in M and ferromagnetic $T_c$ when the layer thickness for the manganites in the composite is decreased from a maximum of 200 μm to a minimum of 10 μm. A similar effect is observed when the sintering temperature exceeded 1000 C [4]. Thus the interface diffusion is a serious problem that needs to be resolved in manganite-PZT structures.

### *3.4 Ferromagnetic resonance*

Investigation of ferromagnetic resonance (FMR) in composites was carried out using a spectrometer operating at 9.8 GHz. The derivative of the microwave absorption line as a function of the magnetic field was registered. Data were obtained for two orientations for H: parallel to the sample plane and perpendicular to the plane. Figure 3 shows absorption versus field profiles for all the samples for H = 0 – 6 kOe. The resonance for NFO-PZT for the field perpendicular to the sample occurs outside the H-range and is not shown in the figure. The resonance field values and the peak-to-peak line widths are given in Table 3.

In general, the resonance field and the line shape depend on many factors, such as the orientation of H, sample geometry, saturation magnetization and anisotropy of the material.

Consider first the resonance fields for parallel and perpendicular orientation of the static field, $H_{//}$ and $H_{\perp}$, respectively. The field $H_{//}$ is expected to decrease and $H_{\perp}$ is expected to increase with increasing magnetization for the samples. The observed $H_{//}$ values, relatively low for NFO-PZT and quite high for LCMO-PZT is then due to a large magnetization for NFO-PZT compared to LCMO-PZT, as discussed earlier with reference to the data of Fig.2. The resonance field $H_{//}$ for LSMO-PZT is low although the magnetization is small. Such a low resonance is indicative of a very high growth induced in-plane magnetic anisotropy during the fabrication of the LSMO-PZT structure. The FMR resonance line width is strongly influenced by the anisotropy of the material. It is seen in Fig.3 that CFO-PZT has a very wide and irregular absorption line, while NFO-PZT has rather narrow and symmetrical lines. This correlates with the hysteresis and coercivity in magnetization for CFO-PZT and absence of it for the NFO-PZT.

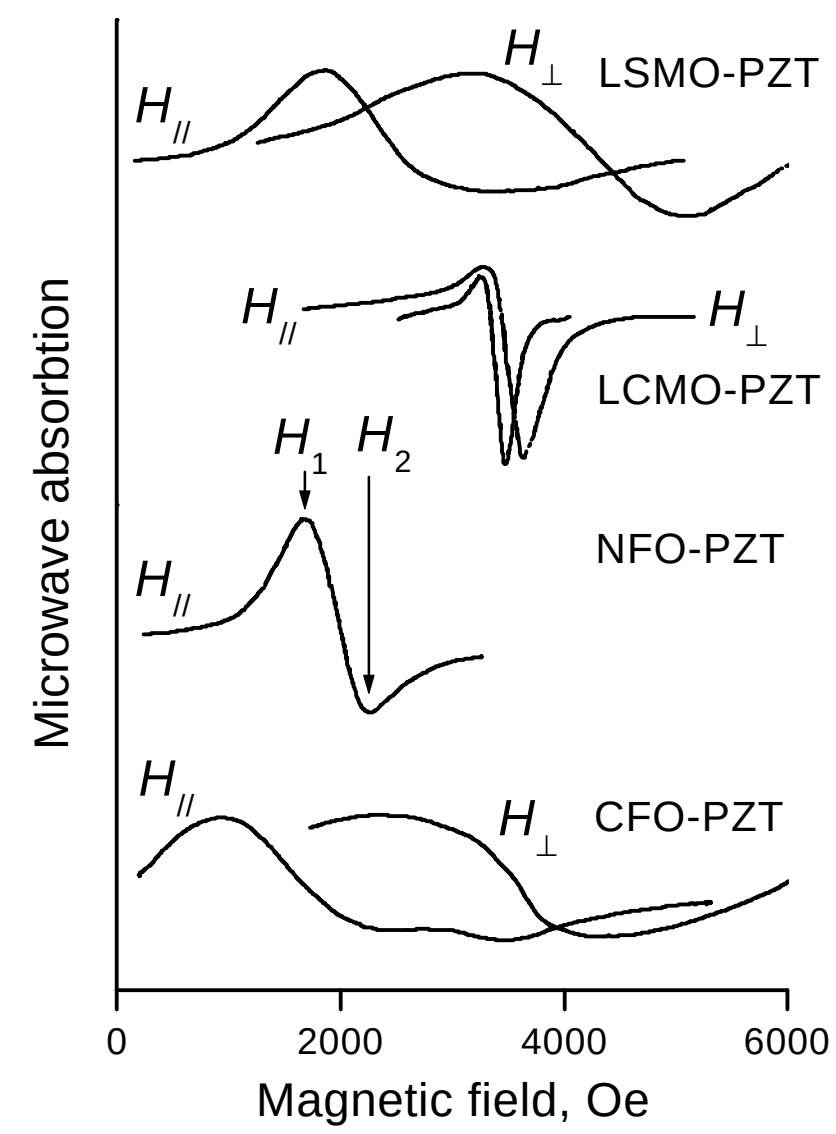

Fig.3: Ferromagnetic resonance spectra for the multiplayer composite structures measured for a frequency of 9.8 GHz at room temperature. Curves denoted "$H_{\perp}$" correspond to static field perpendicular to the sample plane and "$H_{//}$" corresponds to in-plane static fields.

**Table 3**. Ferromagnetic resonance parameters in composite structures at room temperature and 9.8 GHz, and estimated g-value, effective magnetization and anisotropy field.

| Sample | NFO-PZT | CFO-PZT | LCMO-PZT | LSMO-PZT |
|---|---|---|---|---|
| $H_{//}$ | 1.84 | 1.56 | 3.36 | 2.53 |
| $H_{\perp}$ | 6.85 | 3.48 | 3.48 | 4.3 |
| *DH//* | 500 | 1800 | 230 | 1350 |
| $D\,H_{\perp}$ | 400 | 2300 | 640 | 1850 |
| *g* | 2.19 | 3.27 | 2.07 | 2.25 |
| $4pM_{eff}\,(kG)$ | 3.65 | 1.34 | 0.1 | 1.2 |
| $H_A$ (kOe) | 0.5 | 3.7 | 0.15 | -1.1 |

The data on the resonance fields were used to estimate magnetic parameters for the composite. We determined the g-value and the effective magnetization $4\pi M_{eff} = 4\pi M - H_A$, where $H_A$ is the anisotropy field, from H// and $H_{\perp}$, and are given in Table 3. The magnetization data in Fig.2 was then used to calculate the anisotropy field. For NFO-PZT, the g-value and the anisotropy fields are in excellent agreement with values for bulk NFO samples [13]. Thus the two phases in NFO-PZT cosinter very well with practically no diffusion across the interface. For CFO-PZT layered samples, the g-value is higher than the expected value of 2.7 for bulk CFO, but the anisotropy field agrees well with the expected value of 4.3 kOe [13]. It is therefore clear from magnetization and FMR studies that magnetic parameters for the ferrite phase in the layered samples were not affected by high temperature processing. Further, any interface strain also does not appear to have any influence on the anisotropy field.

The samples with manganites, however, show evidence for diffusion of metal ions at the interface. In particular, FMR data reveal a large in-plane anisotropy for LSMO-PZT samples. Since bulk LSMO is magnetically isotropic at room temperature, $H_A$ in the layered sample most likely originates from impurities and/or strain at the interface. The observation is in agreement with a reduction in both the ferromagnetic Curie temperature and magnetization for LSMO-PZT, as discussed in the previous section. The g-values for the layered samples are in agreement with values for bulk manganites [17].

### *3.5 Magnetoelectric effects*

Samples were polished, electrical contacts were made with silver paint, and poled. The poling procedure involved heating the sample to 420 K and the application of an electric field E of 20 kV/cm. As the sample was cooled to 300 K, E was increased progressively to 50 kV/cm over duration of 30 min. The parameter of importance for the multilayers is the magnetoelectric voltage coefficient $\alpha_E$. Magnetoelectric measurements are usually performed under two distinctly different conditions: (i) the induced magnetization is measured for an applied electric effect or (ii) the induced polarization is obtained for an applied magnetic field. We measured the electric field produced by an alternating magnetic field applied to the composite. A set up in which the sample was subjected to a bias field H (with the use of an electromagnet) and an ac field $\delta H$ (1 Oe at 10 Hz – 1 kHz) produced by a pair of Helmholtz coils was used. The sample was shielded from stray electric fields. We implemented lock-in-detection for accurate determination of $\alpha_E$ for two different field orientations: (i) transverse $\alpha_E$ or $\alpha_{E,T}$ for H and $\delta H$ parallel to each other and to sample plane, and $\delta E$ measured perpendicular to the sample plane and (ii) longitudinal or $\alpha_{E,L}$ for all the three fields parallel to each other and perpendicular to the sample plane.

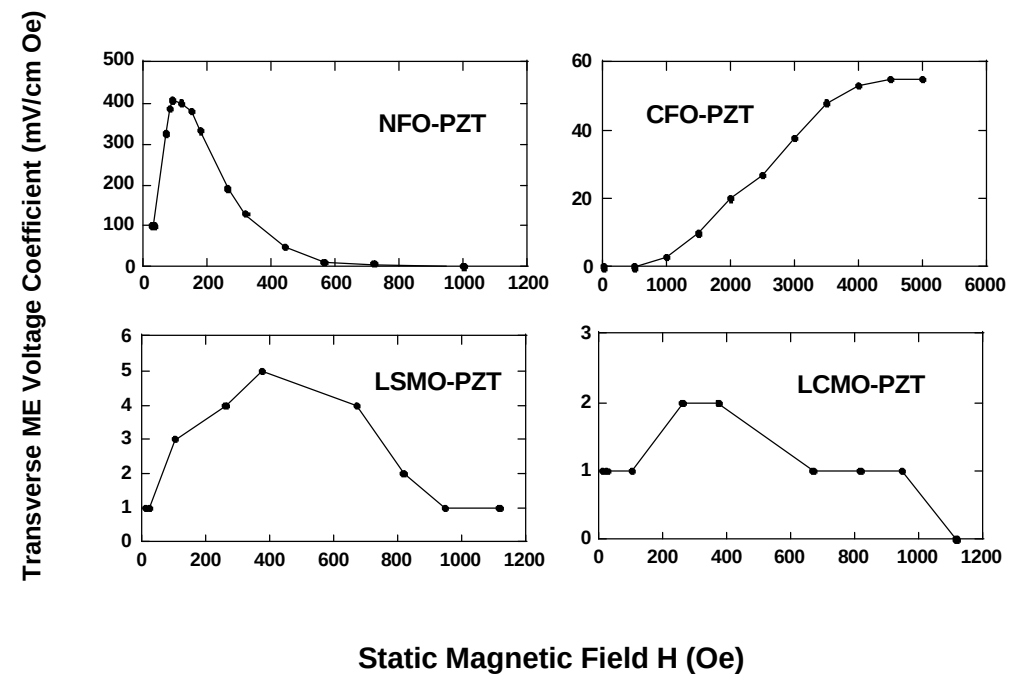

Fig.4: The static field dependence of the transverse magnetoelectric coefficient for multilayers of NFO-PZT, CFO-PZT, LSMO-PZT and LCMO-PZT at room temperature and a frequency of 100 Hz.

The voltage coefficients were measured as a function of bias magnetic field and frequency. Table 4 provides values of the maximum voltage coefficients at room temperature and 100 Hz for all the samples. The values in the table are for unit thickness of the piezoelectric phase. The longitudinal coefficients are smaller by at least an order of magnitude than $\alpha_{E,T}$. Figure 4 shows the static magnetic field dependence of the transverse ME coefficient for the composites at 300 K and 100 Hz ac magnetic field. Consider first the data for NFO-PZT. As H is increased from zero, $\alpha_{E,T}$ increases, reaches a maximum value of 410 mV/cm Oe at $H_{max}$ = 120 Oe and then drops rapidly to zero above 600 Oe. We observed a phase difference of 180 degrees between the induced voltages for +H and -H. As discussed later, the magnitude and the field dependence are related to the slope of magnetostriction vs. H characteristics and can be understood in terms of pseudo-piezomagnetic effects in nickel ferrite. Similar $\alpha_{E,T}$ vs H data were obtained for CFO-PZT multilayers and are shown in Fig.4. The data on $\alpha_{E,T}$ shows an increases with H to a maximum value of 55 mV/cm Oe at H = 5 kOe. The ME effect is present over a wide H-range compared to NFO-PZT. Data for H>5 kOe showed a decrease in $\alpha_{E,T}$.

**Table 4.** Peak transverse and longitudinal magnetoelectric voltage coefficients, $\alpha_{E,T}$ and $\alpha_{E,T,}$ respectively, for the layered samples at room temperature.

| Sample | $a_{E,T}$ *(mV/cmOe)* | $a_{E,L}$ *(mV/cmOe)* |
|---|---|---|
| NFO-PZT | 410 | 45 |
| CFO-PZT | 55 | 5 |
| LSMO-PZT | 5 | 1 |
| LCMO-PZT | 2 | 0.5 |

We observe a rather weak ME effect at room temperature in composites of $La_{0.7}Sr_{0.3}MnO_3$ (LSMO)–PZT and $La_{0.7}Ca_{0.3}MnO_3$ (LCMO)- PZT. Data in Fig.4 indicate orders of magnitude smaller $\alpha_{E,T}$ in manganites-PZT than in ferrite-PZT composites. The effect is stronger in LSMO-PZT than in LCMO-PZT. These observations could be attributed to LSMO layers having a ferromagnetic $T_c$ of 310 K and LCMO being paramagnetic at room temperature. Thus the magnetostriction and the resulting piezomagnetic coefficients for the manganites are expected to be quite small for composites with manganites, leading to poor ME couplings.

Data on the magnetostriction $\lambda$ are necessary for an understanding of the strength of $\alpha_E$ and its H dependence in the composites. Under appropriate bias field H and an ac field $\delta$H, dynamic mechanical strain in the ferromagnetic phase due to magnetostriction gives rise to a pseudo piezomagnetic coupling that facilitates the ME effect. In a layered structure under free body conditions, $\alpha_{E,T}$ is proportional to $\lambda_{//}$, the magnetostriction measured parallel to the in-plane H, and $\alpha_{E,L}$ is proportional to the magnetostriction $\lambda_{\perp}$ measured in the sample plane for H perpendicular to the sample. Figure 5 shows H dependence of both $\lambda_{//}$ and $\lambda_{\perp}$ measured at room temperature with a strain gage and a strain indicator for the ferrites and LSMO samples. The measurements had to be made on bulk samples since the multilayers showed degradation under the influence of epoxy used to cement the strain gage to the sample surface. Measurements on LCMO indicated no magnetostriction since it is paramagnetic at room temperature.

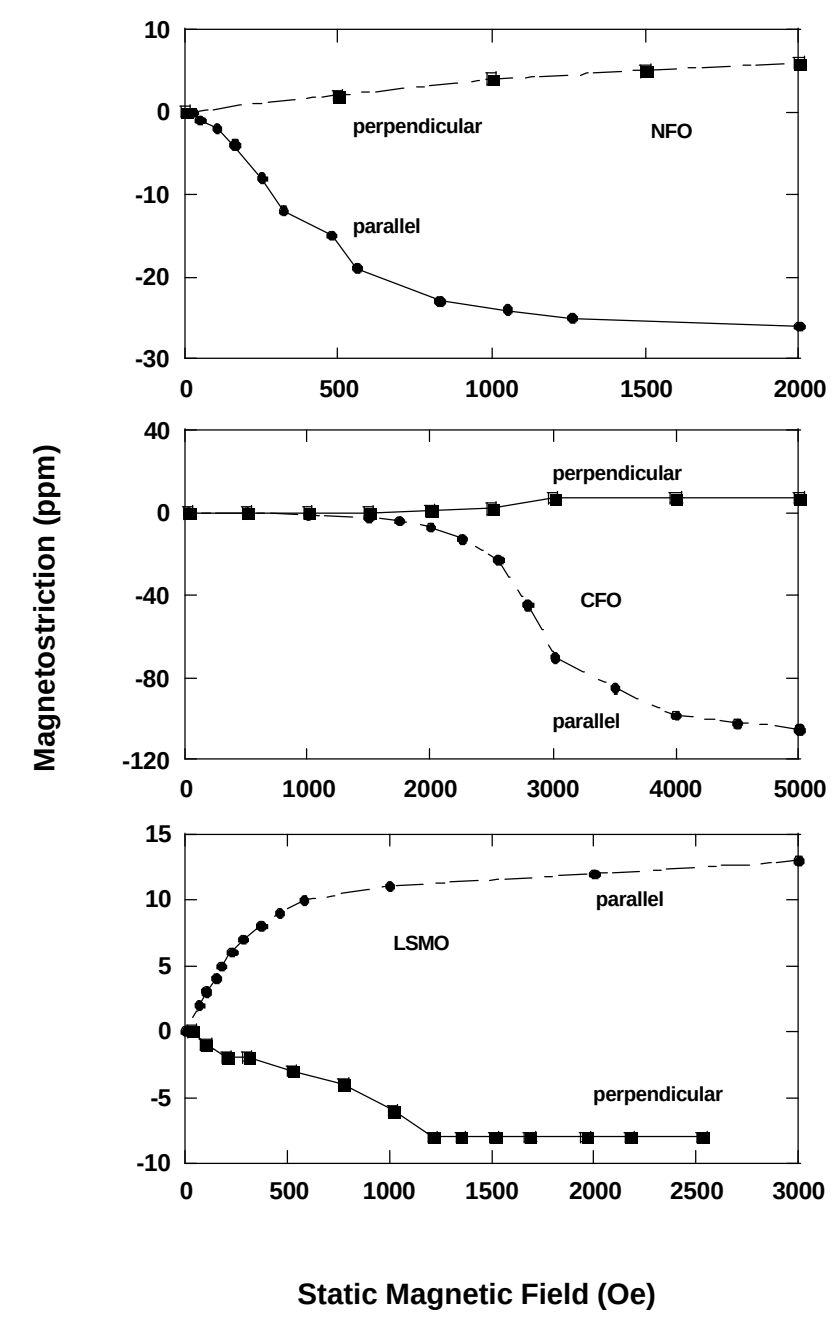

Fig.5: Magnetostriction $\lambda$ as a function of H for bulk samples of NFO, CFO, and LSMO at room temperature. The data are for $\lambda$ measured parallel to in-plane H (parallel) and $\lambda$ measured in-plane for H perpendicular (perpendicular) to the sample plane.

The magnetostriction in Fig.5 increases in magnitude with H for all the samples and levels off at high fields. It is the largest for CFO and the smallest for LSMO. One observes a negative $\lambda_{//}$ for the ferrites while it is positive for LSMO. The saturation value for $\lambda_{//}$ is much higher than for $\lambda_{\perp}$. Since the strength of the piezomagnetic coupling q = $\delta\lambda/\delta H$ is proportional to the slope of $\lambda$ vs H, we anticipate (i) $\alpha_{E,}$ vs H to track the slope of $\lambda$ vs H, (ii) samples of CFO-PZT to show the highest ME voltage coefficients, (iii) a strong ME coupling in NFO-PZT, (iv) the weakest ME effects in LSMO-PZT and (v) an order of magnitude difference in the values of $\alpha_{E,T}$ and $\alpha_{E,L}$ for the ferrites-PZT samples.

Consider first the ME effects in the composites of manganites – PZT. Since LCMO is paramagnetic at room temperature, the piezomagnetic coupling is expected to be small leading to the weakest ME coupling amongst the samples studied here. The data in Fig.4 do reveal $\alpha_{E,}$ values on the order of 1- 2 mV/cm Oe for the samples of LCMO-PZT. The coupling, however, is equally weak in LSMO-PZT samples even though one expects strong ME effects based on $\lambda$ data in Fig. 5. Recall that the magnetostriction is for a bulk sample of LSMO and that magnetization and ferromagnetic resonance (Figs.2 and 3) data imply deterioration of magnetic parameters in the layered composite and consequent weakening of ME coupling. The most likely cause is the diffusion of metal ions at the interface. This observation is in agreement with our earlier studies on bilayers and multilayers of manganite-PZT. We observed the highest value for the ME voltage coefficient in 200 micron thick bilayers of LSMO-PZT. The effect weakened when the layer thickness was increased or the number of layers was increased, observations that point to interface diffusion as the likely cause of weak ME effects in multilayers of manganite-PZT.

Consider next the ME effects in ferrite-PZT composites. In spite of a larger magnetostriction in CFO compared to NFO, Fig.4 shows an order of magnitude higher $\alpha_{E,T}$ in NFO-PZT than in CFO-PZT. Possible reason for enhanced ME effects in samples with nickel ferrite is as follows. There are two types of magnetostriction in a ferromagnet: (i) Joule magnetostriction associated with domain movements and (ii) volume magnetostriction associated with magnetic phase change. The volume magnetostriction is not important in the present situation since it is significant only at temperatures close to the Curie temperature and for high fields. In ferrites domains are spontaneously deformed in the magnetization direction. Under the influence of a bias field H and ac field $\delta H$, both the growth in the domains with favorable orientation and domain rotation contribute to the Joule magnetostriction. Since the ME coupling involves dynamic magnetoelastic coupling, key requirements for the ferrite phase are unimpeded domain wall motion, domain rotation and a large $\lambda$. A soft, high initial permeability (low anisotropy) ferrite, such as NFO, is the main ingredient for strong ME effects. In magnetically hard cobalt ferrite, however, one has the disadvantage of a large anisotropy field that limits domain rotation.

The magnetoelectric coefficient $\alpha_E$ in a composite structure is, therefore, expected to increase with the magnetic permeability of the ferromagnetic phase $\mu=1+4\pi\cdot(\partial M/\partial H)$. It follows from the data of Fig.2 that the highest magnetic permeability $\mu\sim19$ is measured for the NFO-PZT sample. Maximum permeability for the CFO-PZT sample is $\mu\sim3$ and the low permeability is due to influence of the anisotropy as evident from FMR results in Table 3. In both manganite samples with low saturation magnetization, the permeability did not exceed a value of 1.2. These data demonstrate that high magnetic permeability should be realized in multilayer structures to increase the magnetoelectric coefficient. Thus one can relate the strong ME effects in NFO-PZT and poor ME coupling in CFO-PZT to the initial permeability $\mu_i$. In our recent studies on composites with Zn substituted CFO and PZT, the anisotropy decreases and $\mu_i$ increases with Zn substitution. We found excellent correlation between the initial permeability and the strength of ME coupling in the composites.

**4. Conclusion**

Structural, magnetic and magnetoelectric characterization of multilayer composites of ferrite – PZT and ferromagnetic manganite - PZT at room temperature reveal the following. (i) Ferrite-PZT structures are free of any impurities. (ii) Magnetic parameters for ferrite-PZT samples are in agreement with values expected for the ferrite. (iii) Although structural studies do not indicate any impurity phases in manganite-PZT, deviation of magnetic parameters from values for pure manganites implies diffusion of metal ions at the interface. Samples of LSMO-PZT, in particular, show a large magnetic anisotropy. (iv) Ferrite-PZT samples show evidence for strong ME coupling and the effect is quite weak in manganite –PZT. (v) The transverse ME coupling is an order of magnitude stronger than the longitudinal coupling. (vi) The strength of ME coupling correlates with the magnetic permeability.

**Acknowledgments**

The research is supported by a grant from the National Science Foundation (DMR-0072144).